# Методы межкомпонентного взаимодействия в объектно-ориентированных каркасах приложений

В.С.Петросян

Институт проблем информатики и автоматизации

## Резюме

Для осуществления межкомпонентного взаимодействия, т.е. обмена событиями между графическими компонентами внутри приложения, в существующей теории и практике, каркасы используют событийную модель управления, предлагаемую самой операционной системой. В настоящей статье предлагается модель, позволяющая значительно увеличить скорость межкомпонентного взаимодействия за счет использования механизма прямой пересылки сообщений между компонентами, без обращения к операционной системе. Предлагается также механизм подписки, освобождающий от получения ненужных сообщений, что повышает эффективность работы приложений в целом.

## Введение

На сегодняшний день разработка интерактивных графических интерфейсов уже представляется немыслимой без использования каркасов приложений. Каркасы приложений, предназначенные для разработки интерактивных графических интерфейсов, представляют собой набор классов или библиотек, реализующих определенную структуру приложения для конкретной операционной системы. Имея огромное количество реализованных компонентов, допускающих многократное использование, они экономят время программиста, который в противном случае вынужден был бы переписывать большой объем стандартного кода для каждого нового разрабатываемого приложения. В связи с появлением графических пользовательских интерфейсов каркасы приложений получили большое распространение, так как они все определяют стандартную структуру собственных приложений. Следует заметить, что использование стандартного каркаса приложений дает возможность с легкостью создавать инструментарии для автоматизации проектирования графических пользовательских интерфейсов, так как заведомо определена структура кода приложения. В объектно-ориентированных каркасах приложений специфичные части приложения реализуются (имплантируются в систему) путем наследования от существующих в каркасе классов.

**Событийная модель каркасов приложений**

В отличие от традиционных С (консольных) приложений, в которых программа сама управляет процессом выполнения, практически все графические пользовательские системы управляются событиями (event-driven). Это значит, что приложение ожидает события либо инициированного пользователем, либо пришедшего от какого-то другого источника, после чего, переадресовав полученное сообщение соответствующему обработчику, вновь возобновляет режим ожидания.

Таким образом, модель работы каркаса приложений состоит из приема и обработки всех сообщений приложения и переадресации их в случае необходимости соответствующим компонентам.

Межкомпонентное взаимодействие, т.е. обмен событиями между графическими компонентами приложения, каркасы осуществляют используя событийную модель управления, предлагаемую самой операционной системой. Применительно к операционной системе Windows, вследствие того, что каждый графический компонент имеет зарегистрированный за собой класс окна, обращение к нему происходит через дескриптор окна, что в значительной степени замедляет межкомпонентное взаимодействие (рис. 1).

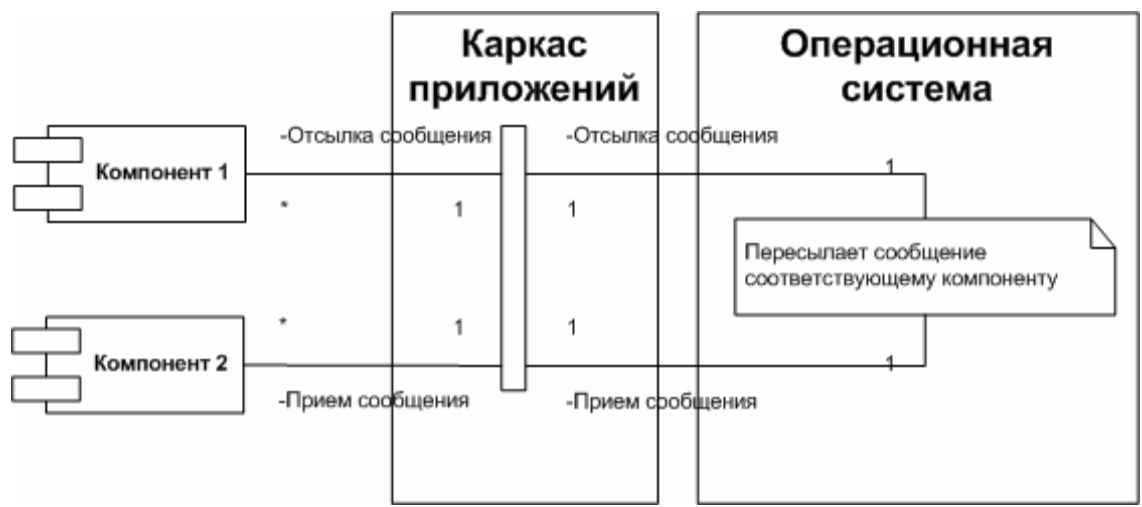

Рис. 1. Межкомпонентное взаимодействие с использованием событийной модели операционной системы

В работе [1] описывался каркас для моделирования взаимодействия игровых компонентов, входящих в состав игровой сцены. Рассмотрим эту модель более обобщенно с точки зрения взаимодействия компонентов внутри каркаса.

Предлагается модель, позволяющая значительно увеличить скорость межкомпонентного взаимодействия за счет использования механизма *прямой пересылки* сообщений между компонентами и механизма подписки на те или иные события (рис. 2).

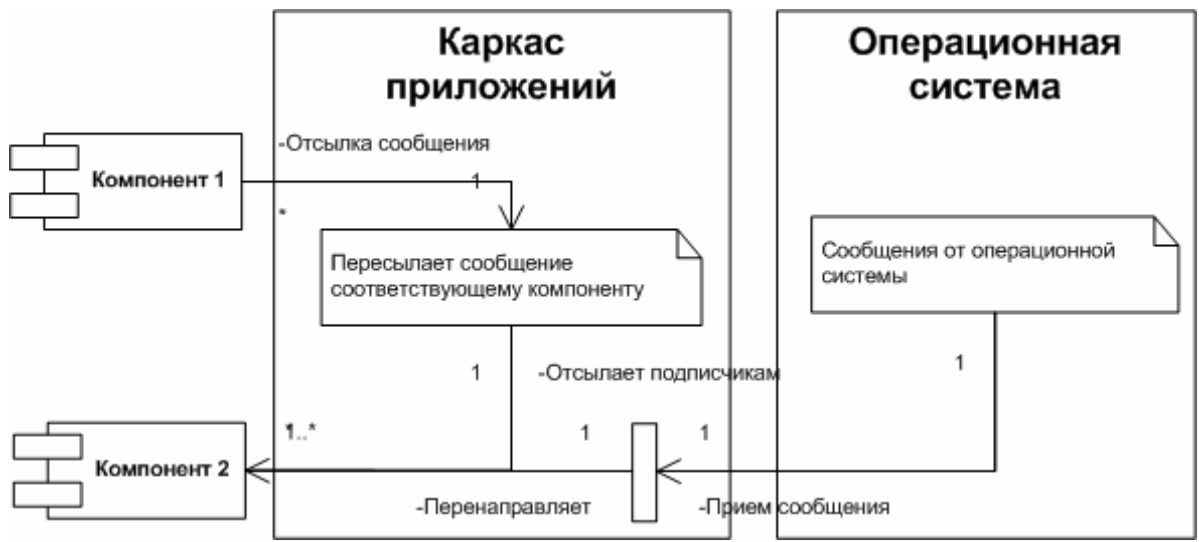

Рис. 2. Межкомпонетное взаимодействие с использованием прямой пересылки.

Следует отметить, что при этом вводится особый тип объектов - *обработчиков событий*, которые могут получать сообщения (например, *компонент 2,* см. рис. 2). При этом, для того чтобы следить за состоянием компонента, обработчики событий становятся подписчиками сообщений компонента, т. е. некоторые объекты регистрируются как объекты обратного вызова, которые должны быть оповещены о совершении того или иного события.

В соответствии с предлагаемой моделью компонент отсылает сообщения не операционной системе, а напрямую обработчикам событий, являющимся его подписчиками.

Поскольку обработчик невольно будет получать сообщения в ответ на все события произошедшие с компонентом, возникает необходимость перераспределения и фильтрации сообщений.

**Обзор существующих моделей**

Рассмотрим ныне существующие модели перераспределения полученных сообщений внутри обработчика.

Разные каркасы предлагают разные механизмы обработки событий. Некоторые объектно-ориентированные каркасы приложений, такие как например Java™ AWT[2], используют метод переопределения виртуальных функций. В таких моделях базовые классы имеют предопределенные типизированные обработчики сообщений, и для обработки того или иного события необходимо как наследование от поставляемого каркасом соответствующего класса обработчика событий, так и переопределение функции обработчика (рис. 3). Однако при этом всевозможные обработчики событий описываются в базовом классе, что является не столь эффективным и сильно влияет на производительность системы в целом. К тому же, это ограничивает круг обрабатываемых сообщений, так как является невыполнимой задачей применительно ко всем типам сообщений.

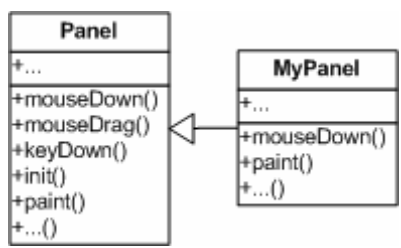

Рис. 3. Переопределение поведения объекта путем наследования

Альтернативным решением является организация *статической привязки* [3,4,5] обработчиков событий к сообщениям благодаря каскадным картам сообщений (например MFC[3,4], Qt[5]). В такой модели каждый класс, имеющий возможность принимать сообщения, имеет свою собственную "карту сообщений", которая используется каркасом для связки полученного сообщения с функцией ее обработки (рис. 4). Следует особо отметить, что каждый класс имеет одну единую карту сообщений для всех своих экземпляров. Однако это не решает вопроса о межкомпонентном взаимодействии, относясь только лишь к распределению и пересылке сообщений к соответствующим обработчикам. При этом межкомпонентное взаимодействие осуществляется путем использования событийной модели операционной системы. Например, использование каркасом MFC событийной модели операционной системы Windows накладывает ограничения на получателя сообщения, которым может являться только объект окна (window).

Следует отметить, что в случае необходимости переопределения предопределенного поведения объекта необходимы не только наследование от соответствующего класса, но и перерегистрация функции обработчика, т.к. функции обработчика, зарегистрированные в "карте сообщений", не являются виртуальными.

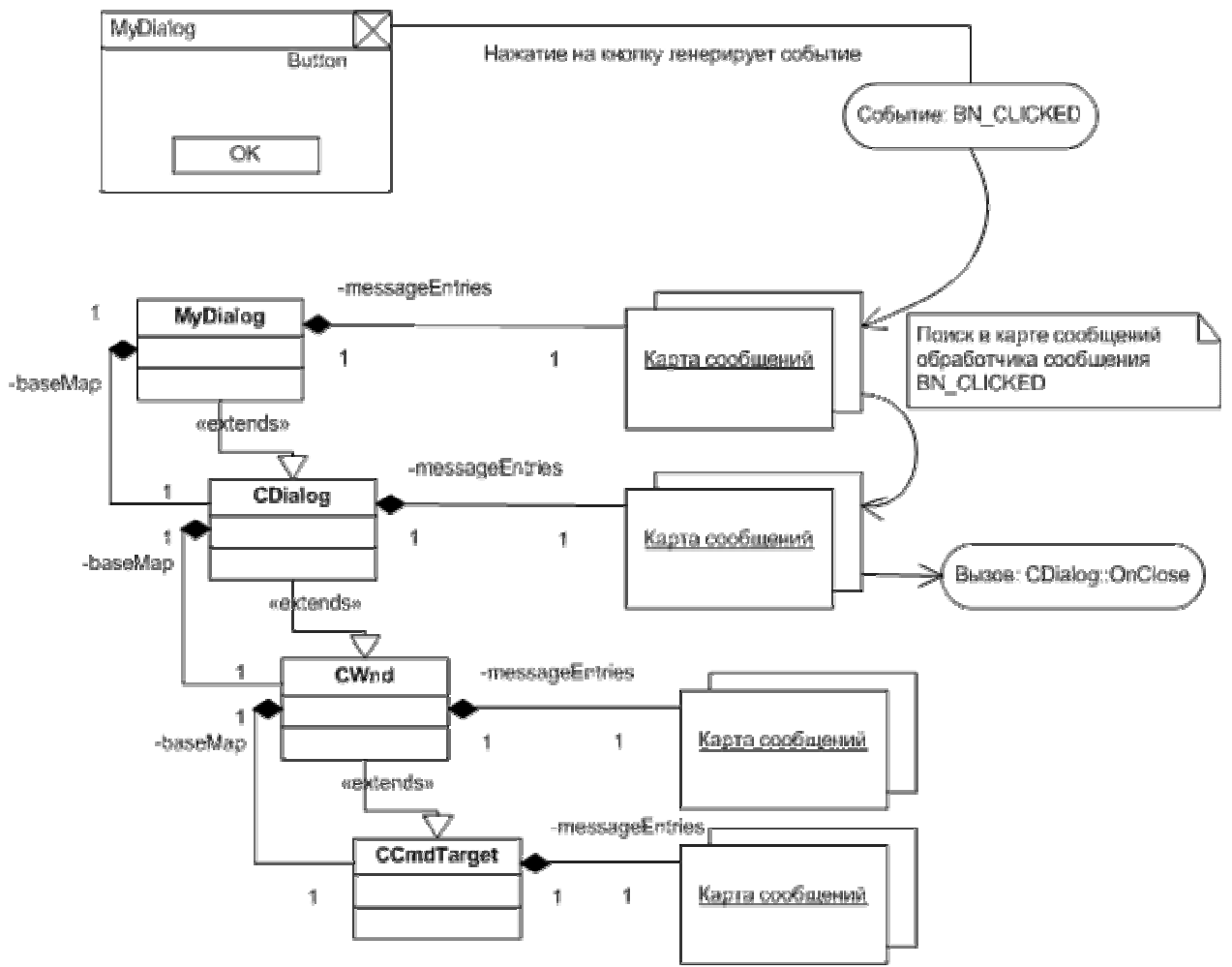

Рис. 4. Модель обработки полученного сообщения в MFC

**Предлагаемая модель**

Рассмотрев вышеописанные сценарии перераспределения полученных сообщений, предлагается следующая модель.

Предлагается ввести особый тип объектов *обработчиков событий,* которые имеют определенный базовый интерфейс для получения и распределения сообщений, а также таблицу обрабатываемых ими сообщений. Экземпляру (instance) данного обработчика доступны только те сообщения о связанных с компонентом событиях, которые зарегистрированны в таблице обрабатываемых им сообщений (рис. 5).

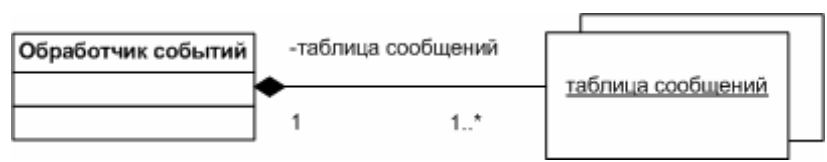

Рис. 5. Таблица входящих сообщений обработчика

Таким образом осуществляется фильтрация ненужных сообщений, и в то же время повышается стойкость системы при возможной модификации компонентов благодаря отсеиванию добавленных позже неизвестных обработчику типов сообщений.

Теперь обратимся к отсылающей сообщения стороне. Если в случае использования событийной модели операционной системы Windows при отсылке сообщения указывается дескриптор окна, которому это сообщение адресовано, а каркасы приложения переадресуют это сообщение соответствующему компоненту обработчика, то в предлагаемой модели отсылающая сторона имеет список подписчиков на сообщения, генерируемые ею, а каркас приложения напрямую передает это сообщение стороне обработчика (рис. 6).

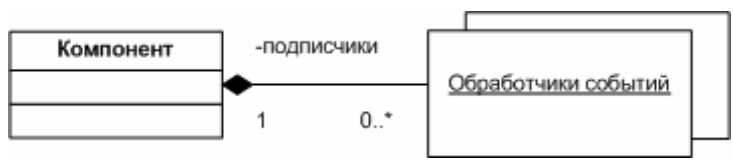

Рис. 6. Компонент содержит список подписчиков

Результаты экспериментов показали, что в случае использования прямой пересылки сообщений достигается значительное увеличение скорости межкомпонентного взаимодействия по сравнению с использованием событийной модели операционной системы Windows.

Ниже приводятся сравнительные результаты теста на отсылку 10 миллионов сообщений от одного компонента графического интерфейса к другому как при событийной модели Windows, так и при предлагаемой модели:

|  | Windows | Предлагаемая модель |
|---|---|---|
| 10 млн отсылок | 39357 мсек | 2814 мсек |
| сообщения/мсек | 254 сообщ/мсек | 3553 сообщ/мсек |
|  |  | **x13.98 быстрее** |

**Заключение**

Таким образом, предлагаемая модель межкомпонентного взаимодействия благодаря прямому обмену сообщений между компонентами приложения обеспечивает повышение скорости осуществляемого межкомпонентного взаимодействия в 13.98 раз по сравнению с использованием стандартных методов, предоставляемых операционной системой Windows. Кроме этого, предлагаемая модель подписки сообщений обеспечивает перенаправление сообщений соответствующим обработчикам, а также их фильтрацию, что позволяет избежать получения ненужных сообщений, повышая эффективность работы приложений в целом.

**Литература**

Петросян Вагинак Суренович,
Институт проблем информатики и автоматизации, аспирант
Tel: +374 10 219763 (сл.), +374 93 2020585 (моб)
E-mail: spl@ipia.sci.am

# Inter-component communication methods in object-oriented frameworks

V.S. Petrosyan

Modern frameworks for development of graphical interfaces are using the native controls of the operating system. Because of that they are using operating system events model for inter-component communication. We consider a method to increase inter-component communication speed by sending messages from one component to the other passing over the operating system. Besides the messages subscription helps to avoid receiving of unnecessary messages.